\documentclass[preprint,authoryear,12pt]{elsarticle}

\usepackage{epsfig}

\usepackage{amssymb}

\usepackage{lscape}

\usepackage[ps2pdf,%
a4paper=true,%
breaklinks=true,%
colorlinks=true,%
pdfauthor={Bianchi \& Garcia},%
pdftitle={UV Spectroscopy of O-type Stars}%
]{hyperref}

\journal{Advances in Space Research}
\newcommand{\Teff}{\mbox{$T_{\rm eff}$}~}
\newcommand{\teff}{\mbox{$T_{\rm eff}$}}
\newcommand{\as}{\mbox{$^{\prime\prime}$~}}
\newcommand{\kms}{km s$^{-1}$}

\newcommand{\Lx}{\mbox{$\log${$L_x$}/$L_{\odot}$}}
\newcommand{\lx}{\mbox{{$L_x$}/$L_{\odot}$}}
\newcommand{\Rstar}{\mbox{$R_{\ast}$}}
\newcommand{\Rsun}{\mbox{$R_{\odot}$}}

\newcommand{\logg}{\mbox{$\log$~\textsl{g}}~}
\newcommand{\vinf}{\mbox{$v_{\infty}$}}
\newcommand{\Vinf}{\mbox{$v_{\infty}$}}
\newcommand{\mdot}{$\dot M$}
\newcommand{\Mdot}{$\dot M$}
\newcommand{\Myr}{M$_{\odot}$ yr$^{-1}$}

\newcommand{\ebv}{$E(B-V)$}

\begin{document}

\begin{frontmatter}



\title{The Effective Temperatures of O-type Stars from UV spectroscopy}

\author[1]{Luciana Bianchi}
\address[1]{The Johns Hopkins University, Dept. of Physics and Astronomy, 3400 N.Charles St.,  Baltimore, MD 21218, USA}
\ead{bianchi@pha.jhu.edu}
\ead[url]{http://dolomiti.pha.jhu.edu}

\author[2]{Miriam Garcia}
\address[2]{
Dept. of Astrophysics,  CSIC-INTA Center for  Astrobiology,
Ctra. Torrejon a Ajalvir km.4, 
28850-Madrid, 
Spain}
\ead{mgg@cab.inta-csic.es}


\begin{abstract}

We present an analysis of high resolution spectra 
in the far-UV $-$ UV range ($\sim$905-2000\AA) with non-LTE,
spherical, hydrodynamical, line-blanketed models, 
of three O-type Galactic stars, and derive their photospheric and wind
parameters.  These  data extend previously analyzed samples
and fill a gap in spectral type coverage. 
The combined sample confirms a  revised (downward) \Teff~ scale 
with respect to canonical calibrations, as found in our previous works from UV and optical spectra, 
and in recent works by other authors.

\end{abstract}


\end{frontmatter}

\parindent=0.5 cm

\section{Introduction}
\label{s_intro}

Massive stars play a crucial role in the chemical and dynamical evolution
of stellar structures in the Universe, for their feedback of energy and 
chemically-enriched material to the interstellar medium. 
Accurate estimates of the physical parameters of massive
stars are relevant for understanding their evolution, and ultimately the chemical 
evolution of galaxies. 
Yet outstanding questions remain to be answered concerning 
 these objects; for 
O-type stars, among the most massive stellar objects, there is still lack of consensus
on their stellar parameters,
such as the effective temperature (\Teff) and mass-loss rate (\mdot) in particular.

In the past decade, significant steps forward were made possible 
 by the \textit{Far Ultraviolet Spectroscopic Explorer}  (FUSE), 
 that enabled access to critical diagnostic lines for hot star winds, 
with spectroscopy in the far-UV ($\Delta\lambda$=905-1187\AA) range.
At the same time, major progress in calculations of stellar atmospheres with 
expanding winds contributed to refined analysis of both UV and optical
diagnostics.
As part of our effort in this field, 
we had 
analyzed far-UV (FUSE),  UV and optical spectra of  
hot massive stars in the Galaxy with
non-LTE model atmospheres with expanding winds, 
to estimate photospheric and wind parameters including 
T$_{eff}$, L$_{bol}$, \logg, \Mdot, and \vinf
(\citet[]{BG02}, \citet[]{GB04} and \citet{BHG09},
hereafter BG02, GB04 and BHG09).
Lines of the most abundant ionic species in the expanding winds, 
including high ionization stages (e.g. OVI) and low-abundance non-CNO 
elements (e.g. SVI, SIV, PV) in the far-UV, 
in addition to NIV, NV, CIV and SiIV transitions
 in the UV range,     play a crucial role for uniquely constraining some 
 stellar parameters. 
Such diagnostics allow us to solve consistently for the 
ionization structure in the stellar wind (e.g., BHG09), and to find 
a solution for the stellar parameters which reconciles lines of different
ionization potential.

The most significant result was that for all stars analysed,
including mid-O types (BG02) and earlier types (GB04),
the \Teff~ derived from our analysis is significantly lower than 
previous estimates, and than values attributed to their
spectral types by classical compilations. Consequently,  
the  luminosities are also  lower. 

Downwards revisions of \Teff were 
independently confirmed by analyses of optical
spectra of hot massive stars with 
state of the art models, calculated with
FASTWIND \citep[]{
Pal05} and CMFGEN \citep[]{HM98,HM99}. Such analyses 
(e.g. \citet[]{MSH05,HPN02,RPH04}), 
also pointed towards lower \Teff's  than previous works
 based on pure H and He models, 
a result qualitatively similar to our findings, 
although in detail small discrepancies remained
among different diagnostics, and modeling treatements. 
Reconciling UV and optical diagnostics can ultimately 
be achieved by an adequate treatement of
clumping in the wind, as shown e.g. by  BHG09 and by Bouret et al. (2012).

Here we present 
FUSE 
observations
of early and mid-O type stars, and a preliminary analysis combining 
these data with UV spectra. 
The paper is arranged as follows:
in Section~\ref{S_RED} the data reduction is explained; in Section~\ref{S_SM}
the spectral morphology of the new stars
together with the previous samples
is described; the quantitative analysis of
the spectra is presented in Section~\ref{S_FIT};
 results are discussed   in Section~\ref{S_DIS}.

\begin{figure}
\label{f_superg_fuse}
\begin{center}
\includegraphics*[width=12.cm]{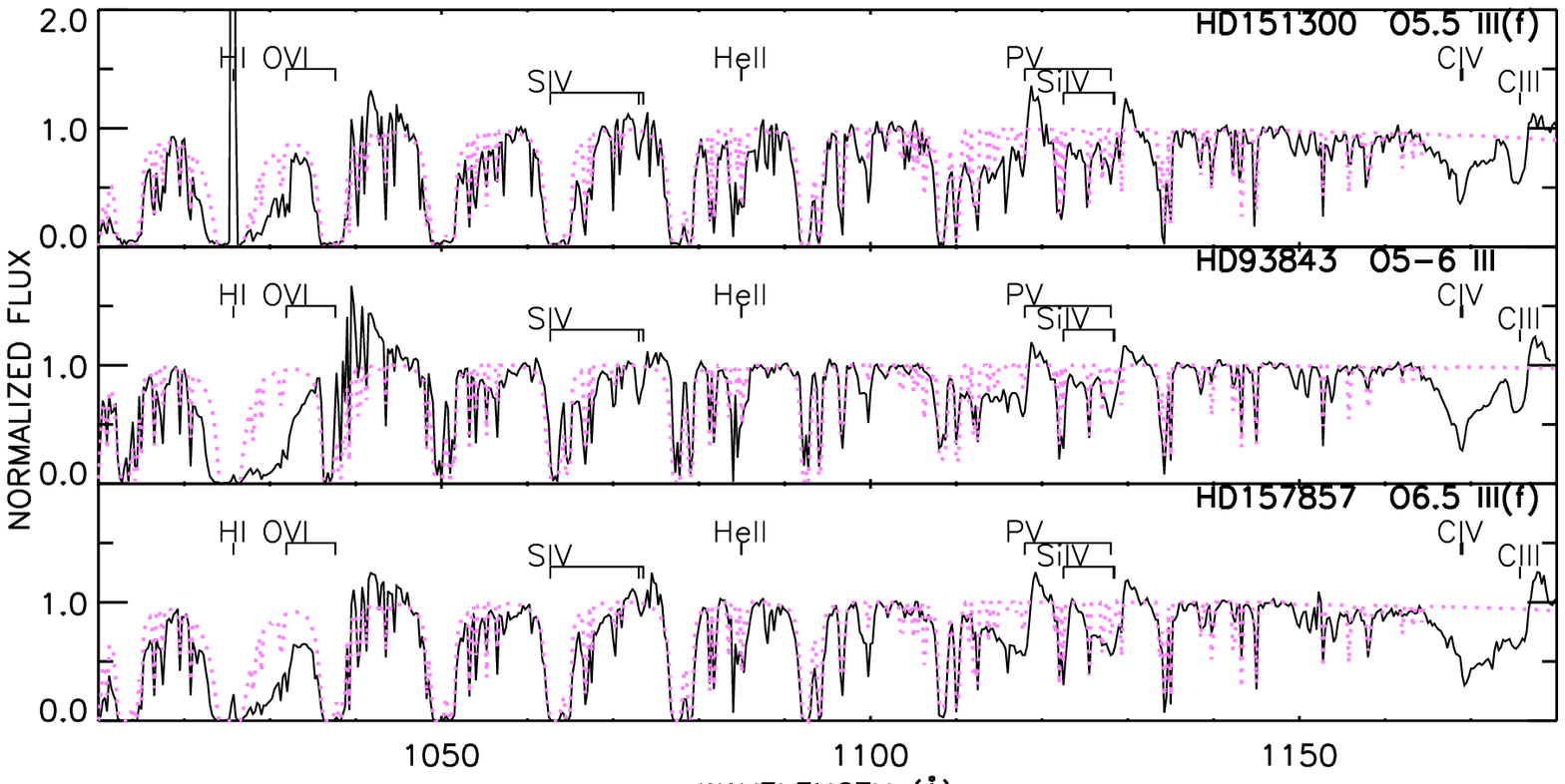}
\vskip -.3cm
\includegraphics*[width=12.cm]{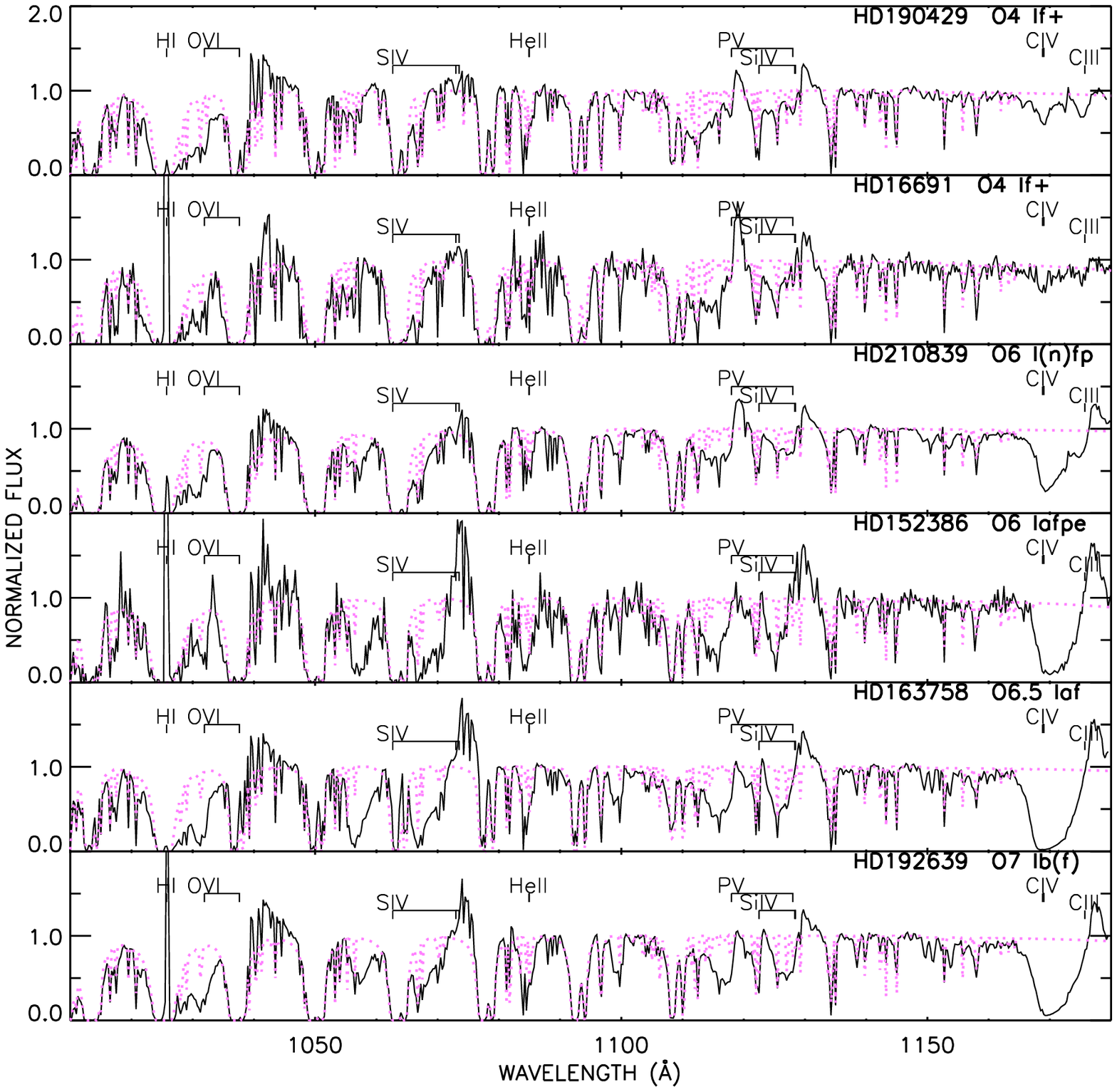} 
\end{center}
\vskip -1.cm 
\caption{{\bf Bottom:}
Far-UV spectra of O-type supergiants.
The pink/grey dotted line is the estimated 
 interstellar hydrogen absorption towards
each object. 
The CIV+CIII blend decreases smoothly towards 
earlier types. PV+SiIV and
SIV peak in intensity around  type
O6 (note, however, that HD~152386 has a ``fe'' type). 
{\bf Top:}
FUSE spectra of HD151300 compared with
the giants analyzed by BG02. 
All lines except OVI are weaker
than in the spectra of the supergiants of similar
spectral type.
 PV+SiIV varies with spectral type 
as in the supergiants sequence.
CIV+CIII decreases towards earlier types.
}
\end{figure}


\section{Spectroscopic Data and Reduction}
\label{S_RED}

\begin{figure}
\label{f_superg_iue}
\begin{center}
\includegraphics*[width=13.cm]{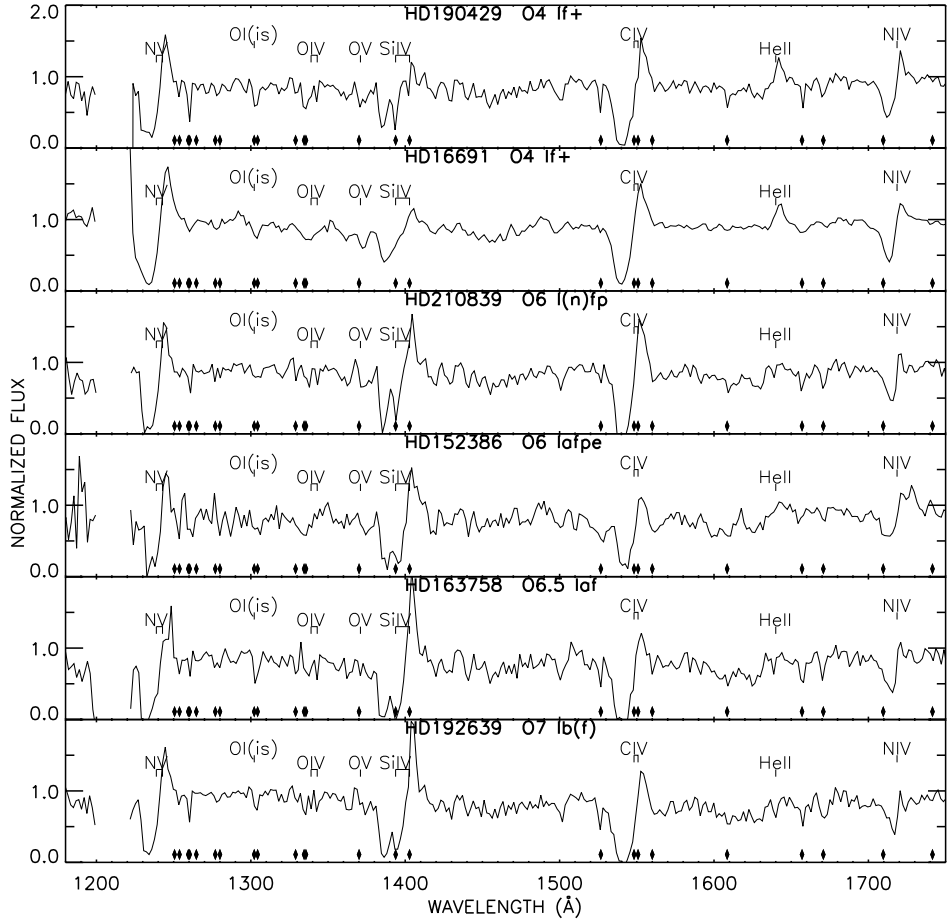} 
\end{center}
\caption{
Spectral morphology of O-supergiants in the UV range.
The diamonds at the bottom of each box mark the most prominent
interstellar lines. 
The strength of
SiIV~$\lambda\lambda$1393.8,1402.8 
decreases from  O7 to  O4 types.
HeII~$\lambda$1640 displays a prominent emission
in the two O4 supergiants.
}
\end{figure}


\begin{table}[]
\caption{Datasets used in this work} 
\begin{tabular}{lcc}
\hline
Star       & \textit{FUSE}     		& \textit{IUE-SWP} \\
\hline
HD~16691   & E8050101			& SWP08323, SWP08288 \\
HD~151300  & E80505-01,-02,-03,-04         & --  \\
HD~152386  & E8050601                      & SWP02848 \\
\hline
\end{tabular}
\label{t_id}
\end{table}

\subsection{FUSE far-UV spectra}

The FUSE data presented here were obtained as part of
program E805 (P.I.: L. Bianchi),
aimed at improving the coverage of the FUSE archive for early
spectral types; in particular, the lack of 
high luminosity objects of the earliest types,
but also the lack of redundancy. The latter is
essential for a meaningful calibration of fundamental stellar parameters, since
the far-UV spectrum of the  most massive stars may reveal peculiarities not predictable
from their optical spectra (e.g. BG02). 
The program was not completed because of FUSE gyros failure; 
 three targets were observed:
HD~16691 (O4~If+, \citet{W73}), HD~151300 (O5.5~III(f), \citet{GHS77})
and HD~152386 (O6~Iafpe, \citet{W73}).

The FUSE data were taken through
the \textit{LWRS} aperture (30\as$\times$30\as)
in \textit{timetag} mode; they
have a resolution of $\lambda/\Delta\lambda \geq $ 20,000.
We aimed at a signal to noise ratio (S/N) of 
S/N$\gtrsim$15 per 0.25\AA~ resolution element; 
one target, HD~151300, was observed for a longer time than planned, 
therefore the S/N of its spectrum is higher.

The data were reduced with the pipeline 
CalFUSE \citep[]{pipeline}. 
All the LiF and SiC segments of every exposure
were examined to verify optimal centering
of the spectra in the extraction window and to avoid
data defects such as event bursts or the ``worm'' 
\citep[]{Sahnow00, Sahnow02}.
To minimize the contamination from airglow lines, only the nighttime
photons were used. We ran the pipeline on individual
exposures, to verify mirror alignment and to correct
for aberrations,  then we combined them and 
subtracted the background. 
In this increased S/N image a more
accurate background subtraction is achieved.

HD~151300 was observed by FUSE during four visits.
We compared the spectra taken in each visit, 
after fully processing them separately, and found
no variations of line profiles or continuum flux  
in visits 2, 3 and 4.
Therefore, to improve S/N, we combined
all raw exposures taken during these visits 
(except for the first exposure of visit 4) before
applying background subtraction.
Visit 1 was not included because it has poorer
S/N; we found differences between the 
flux levels of different segments, and portions
of the spectrum have negative fluxes.
For the other two targets, all exposures and segments were combined.

The spectra were extracted from the combined exposures
and checked with the \textit{IDL} code \textit{fuse\_register}
\citep{L02} for flux consistency and wavelength alignment
among different channels.
Then, the good portions from different channels were 
combined into a final spectrum: SiC2A in the 
region of $\lambda \lesssim $1000\AA, LiF1A in the
1000\AA $ \lesssim \lambda \lesssim $1080\AA~ and
LiF2A for $\lambda \gtrsim $1090\AA. For HD~152386 we used
LiF2B data in the 1000\AA $ \lesssim \lambda \lesssim $1080\AA~
interval because the ``1A'' segments show negative portions
despite the improved background subtraction.
The wavelength ranges 905-930\AA\space and
1181-1187\AA, at the ends of the \textit{FUSE} range,
have poor quality and were not used in this work.
The 1082.5-1087\AA\space region is only covered by SiC channels, 
therefore the HeII~$\lambda$1084.9 line 
was given a small weight in the analysis.

The resulting combined spectra,
normalized to the local continuum,
are shown in Figure~\ref{f_superg_fuse}. 
These plots also show
a model of interstellar atomic and molecular hydrogen absoprtions 
calculated for the line of sight of each star (see
BG02 and GB04 for details)
using \ebv~ values 0.92, 0.71 and 0.80 
for HD~16691, HD~151300 and HD~152386, taken from the literature. 
These $H_{2}+HI$ calculated absorption spectra are useful to assess which 
observed features 
 are purely stellar and to determine reliable
continuum regions for flux normalization.
Interstellar H$_2$ and HI  absorption 
is severe in the FUSE wavelength range. 

\subsection{UV spectra}

IUE observations (1200-3200\AA) exist for HD~152386 (low resolution, 
$\Delta\lambda\approx$ 6\AA) and HD~16691 
(at resolution of both $\Delta\lambda\approx$ 0.2\AA~ and 6\AA).
With the tools in the MAST database, 
we examined all the observations of both targets
taken with the SWP camera (1150-1975\AA),
to check for variability and to exclude saturated portions, and
chose the data with the best S/N.
We discarded the high resolution spectrum of HD~16691
because some portions 
have negative flux.
We then downloaded the selected spectra
from the INES archive \citep[]{ines} because
the data typically have a better background correction 
at wavelengths shorter than 1400\AA, where overlap between echelle orders is 
significant (Bianchi \& Bohlin 1984).
HD~151300 has not been observed in this spectral range.
The normalized IUE spectra in the wavelength range containing the strongest
spectral lines (1200-1750\AA) are shown in Figure \ref{f_superg_iue}.
Table \ref{t_id} compiles the FUSE and IUE
datasets used in this work.

\section{Spectral line morphology}
\label{S_SM}

In Figures \ref{f_superg_fuse} and  \ref{f_superg_iue} we compare
the FUSE (1010-1180\AA~ portion) and IUE (1200-1750\AA)
spectra of HD~16691 and HD~152386 with the spectra of supergiant
stars analysed by BG02 and GB04. 
The combined sample covers spectral types O4 to O7,
and line variations across this range can be examined.

Figure~\ref{f_superg_fuse} shows that the strength of
OVI~$\lambda\lambda$1031.9,1037.6
is approximately constant through spectral types O4-O7. 
SIV~$\lambda\lambda$1062.7,1073.0,1073.5 display
strong P~Cygni profiles in HD~152386 (the emission of the
bluest component being masked by hydrogen absorption)
as in the other mid-O supergiants; 
these lines have maximum strength around
type O6 and  weaken towards both earlier
and later types.

The PV~$\lambda\lambda$1118.0,1128.0 + SiIV~$\lambda\lambda$1122.5,1128.3,1128.4 
blend has  similarly maximum strength 
around type O6 but smaller variation across these types. 
Its  profile changes, with  the absorption being 
deeper on the shorter wavelength side at earlier spectral  types, 
suggesting that these lines are formed in different layers of the wind
\citep[]{Bal00}.  These lines are also known to be sensitive to wind clumping. 
The strength of the CIV$\lambda$1169 + CIII$\lambda$1176 lines,
which we found in BG02 and GB04 to be a temperature indicator 
for stars of similar luminosity class, decreases
from the O7 to the O4 supergiants.

The present sample includes two O6 supergiants, HD~210839
and HD~152386. The lines in the FUSE spectra of HD~210839 are 
generally weaker than in the other O6-O7 supergiants,
as pointed out by BG02.
The line strength in the spectrum of HD~152386 follows
better the smooth trend of variations with spectral type
observed across the sample,
although HD~210839 has the same luminosity class
based on its optical spectra.

The most prominent features in the IUE range
(Figure~\ref{f_superg_iue})
are SiIV~$\lambda\lambda$1393.8, 1402.8, 
NV~$\lambda\lambda$1238.8,1242.8
and CIV~$\lambda\lambda$1548.2,1550.8
which show fully developed P~Cygni profiles in all the supergiants.
SiIV~$\lambda\lambda$1393.8,1402.8 follows the
behaviour of CIV$\lambda$1169 + CIII$\lambda$1176:
it decreases towards earlier types. 
The emission of 
CIV~$\lambda\lambda$1548.2,1550.8
is stronger in the early type supergiants and HD~210839.
NV~$\lambda\lambda$1238.8,1242.8 does not exhibit
a clear trend, in this range. 

The OIV~$\lambda\lambda$1339,1343
and OV~$\lambda$1371 absorptions, 
and the NIV~$\lambda$1718 P~Cygni profile,
grow stronger towards the earlier types. The spectra of HD~190429A 
and HD~16691 (both O4~If$^+$) display 
HeII~$\lambda$1640.0 emission, characteristic of O3-O5~If stars 
\citep[]{IUEatlas,WN87}, which indicates high mass-loss rate (see GB04).
The spectra of the O4 supergiants are very similar to each other,
but NIV~$\lambda$1718, SiIV~$\lambda\lambda$1393.8,1402.8, 
CIV~$\lambda\lambda$1548.2,1550.8 and 
PV~$\lambda\lambda$1118.0, 1128.0 +
SiIV~$\lambda\lambda$1122.5,1128.3,1128.4
are stronger in HD~16691. This can be explained by a higher
mass-loss rate for this star or, as shown in Section~\ref{S_FIT_16691},
a lower effective temperature.

In Figure~\ref{f_superg_fuse} (top) 
the FUSE spectrum of HD~151300
is compared with other mid-O type giants studied
in BG02. All spectral features exhibit weaker profiles than the supergiants in the same figure,
except for OVI~$\lambda\lambda$1031.9,1037.6 
which has similar strength.
The most remarkable case is SIV~$\lambda\lambda$1062.7,1073.0,1073.5,
with almost no wind profile.
Although the comparison covers a smaller spectral range, both 
PV~$\lambda\lambda$1118.0,1128.0 + SiIV~$\lambda\lambda$1122.5,1128.3,1128.4
and CIV$\lambda$1169 + CIII$\lambda$1176
show variations similar to those seen in supergiants.

\section{Analysis of the UV Spectra  with Model Atmospheres}
\label{S_FIT}

\begin{figure}
\label{f_hd151300_152386}
\begin{center}
\includegraphics[width=4.cm,angle=90]{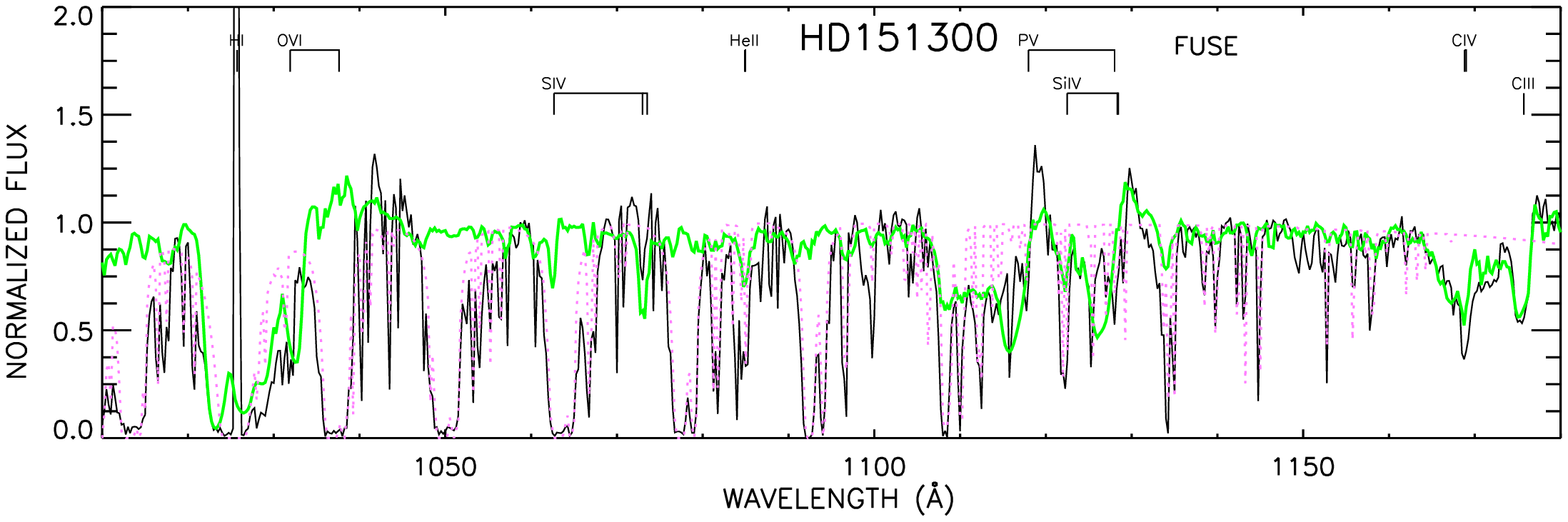}
\includegraphics[width=7.8cm,angle=90]{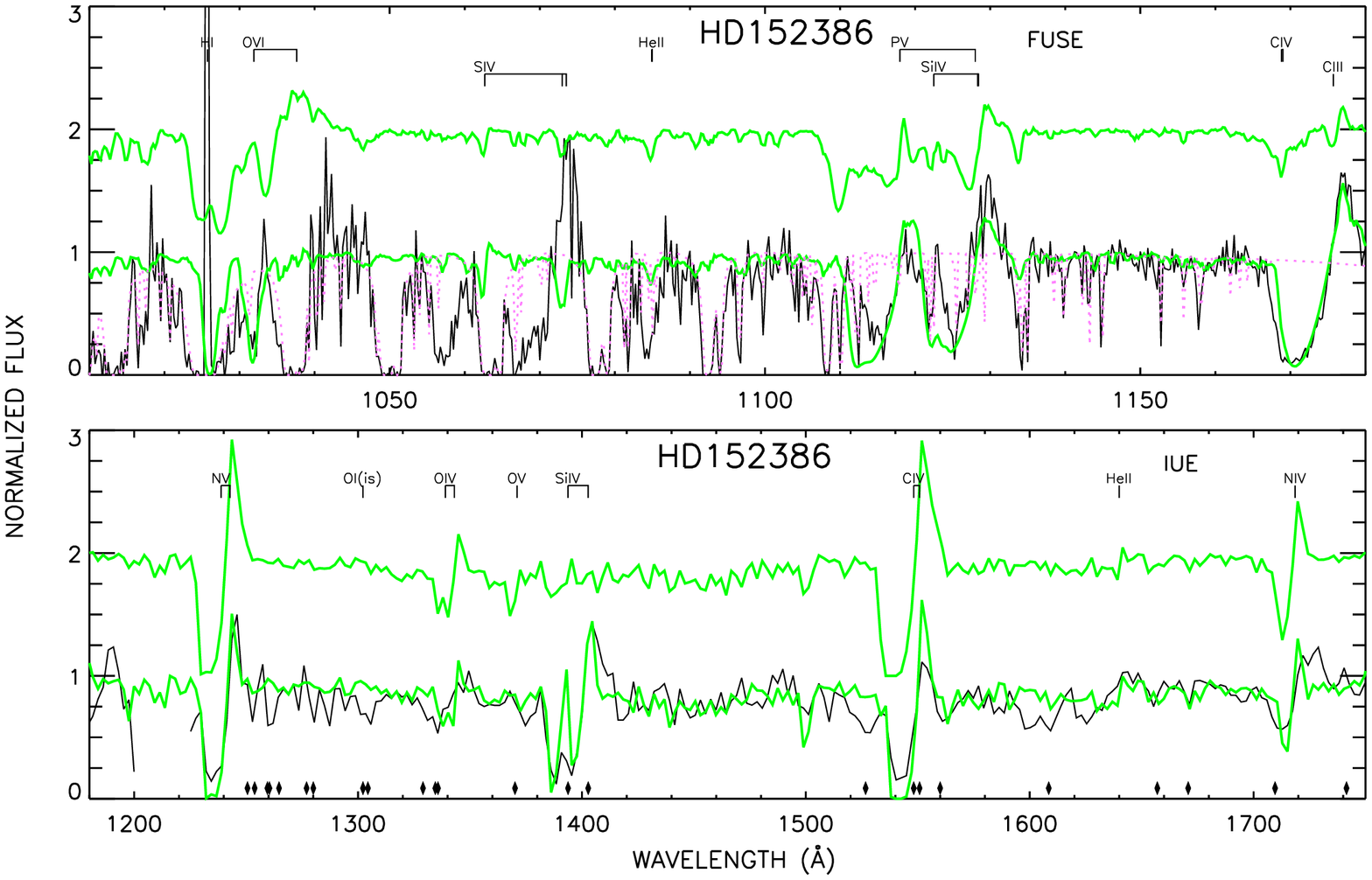}
\end{center}
\caption{{\bf Top:}
FUSE spectrum (black) and 
best fit WM-basic model (green/light grey) of HD~151300.
Interstellar hydrogen absorptions 
(estimated from the reddening value) are shown with a pink/grey-dotted line.
The model parameters are given in Table 2. 
{\bf Bottom:}
 HD~152386 FUSE (top plot) and IUE (bottom plot) spectra. 
The positions of the major interstellar lines in the IUE range are indicated
with diamonds at the bottom.
The model plotted over the observed spectrum is the best-fit model,
with \Teff=34,000~K, \logg=3.3, \Rstar=21\Rsun, \Mdot=$\rm 4.2 \cdot 10^{-6}$ \Myr,
\vinf=1800\kms~ and \Lx=-7.0.
A higher temperature model reproduces better the observed
emission of OVI (upper model, 
 \Teff=38,000~K, \logg=3.4, \Rstar=21\Rsun, \Mdot=$\rm 5.4 \cdot 10^{-6}$ \Myr,
\vinf=2300\kms~ and \Lx=-7.0) however it does not 
match other lines, notably the CIII$\lambda$1176 P~Cygni profile.
}
\end{figure}

We analyzed  the far-UV -- UV spectra of the program stars,
using synthetic atmosphere models computed with the WM-basic code \citep[]{WMBAS}.
WM-basic applies a consistent non-LTE treatment of  line-blocking  
to the entire atmosphere and solves the hydrodynamical structure of the 
spherically expanding envelope, with a smooth transition
from the quasi-static photosphere to the high velocity outflow.
Soft-X rays produced by shocks in the wind, 
which affect the ionization of most stages observed in UV,
are also taken into account.
This parameter can be constrained thanks to the
OVI~$\lambda\lambda$1031.9,1037.6 
doublet in the FUSE range.
The best-fitting models obtained from this analysis are shown in Figures~3 and 4, 
and the derived stellar parameters are listed in Table~\ref{t_models}.
Below we discuss the analysis of the individual stars.

\subsection{HD~151300}

The terminal velocity, \vinf$\simeq$2700 \kms, was estimated from 
v$_{edge}$, measured at the short-wavelength
edge of the OVI~$\lambda\lambda$1031.9,1037.6 absorption, 
reduced by 20\% to account for 
 turbulence \citep{GLP89}. 
We computed a small grid of WM-basic models with this terminal velocity,
\logg=3.6 and \Rstar=16\Rsun~ (typical for giant stars).
\Mdot, \Lx~ and \Teff~ were constrained from
PV~$\lambda\lambda$1118.0, 1128.0 + SiIV~$\lambda\lambda$1122.5, 1128.3, 1128.4
and CIV$\lambda$1169 + CIII$\lambda$1176, whose strength increases 
by increasing \Mdot~ and decreasing \Lx~ and \Teff, in this regime, 
and OVI~$\lambda\lambda$1031.9,1037.6 which has the
opposite dependence on these parameters.

The lower limit of \Teff  is 32,000~K, below which
CIV$\lambda$1169 + CIII$\lambda$1176
displays a saturated absorption in the models,
in contrast with the observed spectrum, for the high \Mdot~ needed to fit
PV~$\lambda\lambda$1118.0,1128.0 + SiIV~$\lambda\lambda$1122.5,1128.3,1128.4.
This behaviour is observed for the \Lx-range where 
OVI~$\lambda\lambda$1031.9,1037.6 is produced (\Lx$\geq$-7.5).
The upper limit for \Teff~ is 38,000~K, when
PV~$\lambda\lambda$1118.0,1128.0+SiIV~$\lambda\lambda$1122.5,1128.3,1128.4
in the models become stronger than observed for the high mass-loss rates needed
to produce any CIV$\lambda$1169+CIII$\lambda$1176.

The best-fit model has \Teff=35,500~K and is shown in
Figure~3. 
The value of  \Lx~ was determined by fitting
OVI~$\lambda\lambda$1031.9,1037.6.
The upper/lower limits for \Mdot~ where
set to the values that produced
PV~$\lambda\lambda$1118.0,1128.0 + SiIV~$\lambda\lambda$1122.5,1128.3,1128.4
and CIV$\lambda$1169 + CIII$\lambda$1176
in excess/defect in the models.
We also varied \logg~ and \Rstar~ until the resulting model
did no longer fit the observations, yielding the error bars for these parameters.

\subsection{HD~152386}

The terminal velocity was determined from $\rm v_{edge}$
measurements of all the spectral transitions showing 
a wind profile. However, none of these lines is saturated,
except for
OVI~$\lambda\lambda$1031.9,1037.6 which is 
not useful since the
bluest part of its absorption overlaps with an H$_2$ transition
(Figure~\ref{f_superg_fuse}). Therefore we can only
obtain a lower limit of \vinf$\gtrsim$1800\kms.

Our extensive grid of WM-basic models 
for supergiant stars (\logg=3.4, \Rstar=21\Rsun)
indicates that 
\Mdot~ $\rm \geq 1 \cdot 10 ^{-6}$ \Myr~
is required to match the observed P~Cygni of 
CIII$\lambda$1176. In this  regime of \mdot,
\Teff  is constrained with 
PV~$\lambda\lambda$1118.0,1128.0 + SiIV~$\lambda\lambda$1122.5,1128.3,1128.4,
CIV$\lambda$1169 + CIII$\lambda$1176 and
SiIV~$\lambda\lambda$1393.8,1402.8 
(the three having the same dependence on \Teff, \Mdot~ and \Lx).
NIV~$\lambda$1718, which increases
with increasing \Teff~ and \Mdot~ but is not sensitive to \Lx,
provides a further constraint.
The lower limit is \Teff=32,000~K.
Models with lower temperatures and the high mass-loss rates
needed to match CIV$\lambda$1169 + CIII$\lambda$1176 absorption,
have stronger than observed
PV~$\lambda\lambda$1118.0,1128.0 +
SiIV~$\lambda\lambda$1122.5,1128.3,1128.4
and  SiIV~$\lambda\lambda$1393.8,1402.8, even when extremely high shocks
(\Lx=-6.3) are assumed. In these models, the emission of 
CIV$\lambda$1169+CIII$\lambda$1176 is not as strong as 
observed and NIV~$\lambda$1718 is too weak.
The upper limit of the effective temperature is 36,000~K. Above this value
CIV$\lambda$1169 + CIII$\lambda$1176 does not develop
a P~Cygni profile and SiIV~$\lambda\lambda$1393.8,1402.8 
is absent, according to our models,  even at high mass-loss rates that lead
to  too strong 
PV~$\lambda\lambda$1118.0,1128.0 + SiIV~$\lambda\lambda$1122.5,1128.3,1128.4.
lines in the model spectra.

\Mdot~ and \Lx~
were constrained from the model that provides the
best overall fit to 
OVI~$\lambda\lambda$1031.9,1037.6,
PV~$\lambda\lambda$1118.0,1128.0 + 
SiIV~$\lambda\lambda$1122.5,1128.3,1128.4,
 CIV$\lambda$1169+CIII$\lambda$1176,
SiIV~$\lambda\lambda$1393.8,1402.8 
and NIV~$\lambda$1718.
We then varied \logg~ and \Rstar~ to find the acceptable
range of values for these parameters,
although they can only be loosely constrained from wind lines.
The best fit model is shown in
Figure~3. 

Models with high mass-loss rates, as required
to fit the CIII$\lambda$1176 lines,
do not match the observed OVI~$\lambda\lambda$1031.9,1037.6 
emission in the \Teff=32,000-36,000~K range.
Higher \Teff's are needed to fit
this doublet: in Figure~3 
we compare the best-fit model with one having similar parameters but 
\Teff=38,000~K. The higher \Teff model matches the
OVI~$\lambda\lambda$1031.9,1037.6  profile, but
PV~$\lambda\lambda$1118.0,1128.0 + SiIV~$\lambda\lambda$1122.5,1128.3,1128.4,
CIV$\lambda$1169+CIII$\lambda$1176 and
SiIV~$\lambda\lambda$1393.8,1402.8 are fainter,
and NIV~$\lambda$1718 and OIV~$\lambda\lambda$1339,1343
stronger than in the observed spectrum, 
indicating that \teff=38,000~K is too high.
The 34,000~K  model does fit the absorption
at $\lambda\sim$1032\AA, which is
the blue-shifted red component of the doublet.
This analysis indicates that a different treatement of shocks in the winds
and inclusion of clumping 
is needed (e.g. BHG09), which we plan as a continuation of this program with the CMFGEN code.

The best-fit model fails to match the strong 
SIV~$\lambda\lambda$1062.7,1073.0,1073.5
profiles and the unsaturated absorption of
NV~$\lambda\lambda$1238.8,1242.
These problems, also encountered in BG02 and GB04, 
probably originate in the sulfur data used by
WM-basic and the treatment of shocks. 
There is also a mismatch between the
model and observed CIV~$\lambda\lambda$1548.2,1550.8 lines.
 While it is possible to fit
their  profile with models of sub-solar carbon abundance
($X _C = 0.01 \;  X _{C,\odot}$), such models  cannot
match the observed
CIV$\lambda$1169 + CIII$\lambda$1176.
HD~152386 is a binary (detected only with speckle interferometry,
\citealt[]{Mal98}) but this fact alone cannot account for the discrepancy,
since a possible contribution by the companion would also affect 
other spectral lines.

\subsection{HD~16691}
\label{S_FIT_16691}

The spectra of this star and our analysis to constrain its parameters are rather similar to 
 HD~190429A, analyzed by  GB04.
However, no consistent fit of all line diagnostics could
be obtained using models with solar abundances.
The best-fit model with solar abundances is shown in  Figure~4, 
upper panel.
In this model, the strength of 
OVI~$\lambda\lambda$1031.9,1037.6, 
CIV~$\lambda\lambda$1548.2,1550.8 
and HeII~$\lambda$1640.0 are discrepant 
with the observations. A much better match is obtained
by using CNO abundances for a slightly evolved
star 
$X _{He} = 5 \; X _{He,\odot}$,
$X _C = 0.5 \;  X _{C,\odot}$,
$X _N = 2.0 \;  X _{N,\odot}$,
$X _O = 0.1 \;  X _{O,\odot}$
(lower panel in Figure~4). 
The analysis thus indicates that this star may be evolved.
The stellar parameters (given for both fits in Table~\ref{t_models}) do not differ
significantly between the models with different abundances, 
therefore the abundance uncertainty does not influence our
\Teff~ calibration.  Bouret et al. (2012) also analyzed this star, using the CMFGEN code, and 
found a slightly higher \Teff (40,000K, {\it vs} our result of \Teff=37,000$\pm$2,000K), but otherwise very similar parameters, and 
C,O sub-solar, N super-solar abundances, similar to what our analysis with WM-basic indicates.


\begin{landscape}
\begin{table*}
\hskip -2.cm
\begin{center}
\footnotesize
\caption{Stellar parameters from model fitting of far-UV and UV spectra. } 
\begin{tabular}{llccccccccc}
\hline
Star                      & Sp.Type (ref)   
           &\Teff         & log~g          & R  & M  & L$_{bol}$/L$_{\odot}$ & \Vinf
           & \mdot                         & \lx          \\
                          &             
           & (kK)         &                &  ((R$_{\odot}$))            &   M$_{\odot}$           &        log                   & (\kms)
           & 10$^{-5}$ \Myr                          &       log       \\
\hline
HD~16691$^a$              & O4~If$^+$(1)  
           &37.0$\pm$2.   & 3.4$\pm$.3    &25$\pm$3   & 57  & 6.0$\pm$0.2    & 2100$\pm$200  &
              1.0$\pm$0.2  &  -6.75$\pm$0.5\\
            &       
           &36.5$\pm$2.   & 3.4$\pm$0.3    &25$\pm$3   & 57  & 6.0$\pm$0.2    & 2200$\pm$200  &
              0.9$\pm$0.2  &  -6.75$\pm$0.5\\
HD~151300  & O5.5~III(f)(2)  
           &35.5$\rm ^{+2.5}_{-3.5}$   & 3.7$\pm$0.3    &16$\pm$3   & 47  & 5.6$\pm$0.3    & 2700$\pm$300  &
              0.2$\pm$0.3  &  -6.5$\pm$0.5\\
HD~152386  & O6~Iafpe (1)  
           &34.0$\pm$2.   & 3.3$\pm$0.3    &21$\pm$3   & 40  & 5.7$\pm$0.2    & $\gtrsim$1800  &
              0.4$\pm$1.0  &  -7.0$\pm$0.5\\
\hline
\end{tabular}
\label{t_models}
\noindent{\small References:(1) \citet[]{W73}, (2) \citet[]{GHS77} \\
$a$~ The first line lists the parameters for the best-fit model with solar abundances,
the second line for the model with CNO abundances: 
$X _{He} = 5 \; X _{He,\odot}$,
$X _C = 0.5 \;  X _{C,\odot}$,
$X _N = 2.0 \;  X _{N,\odot}$,
$X _O = 0.1 \;  X _{O,\odot}$ }
\end{center}
\end{table*}
\end{landscape}

\section{Discussion}
\label{S_DIS}

\begin{figure}
\label{f_hd16691}
\begin{center}
\includegraphics*[width=8.5cm,angle=90]{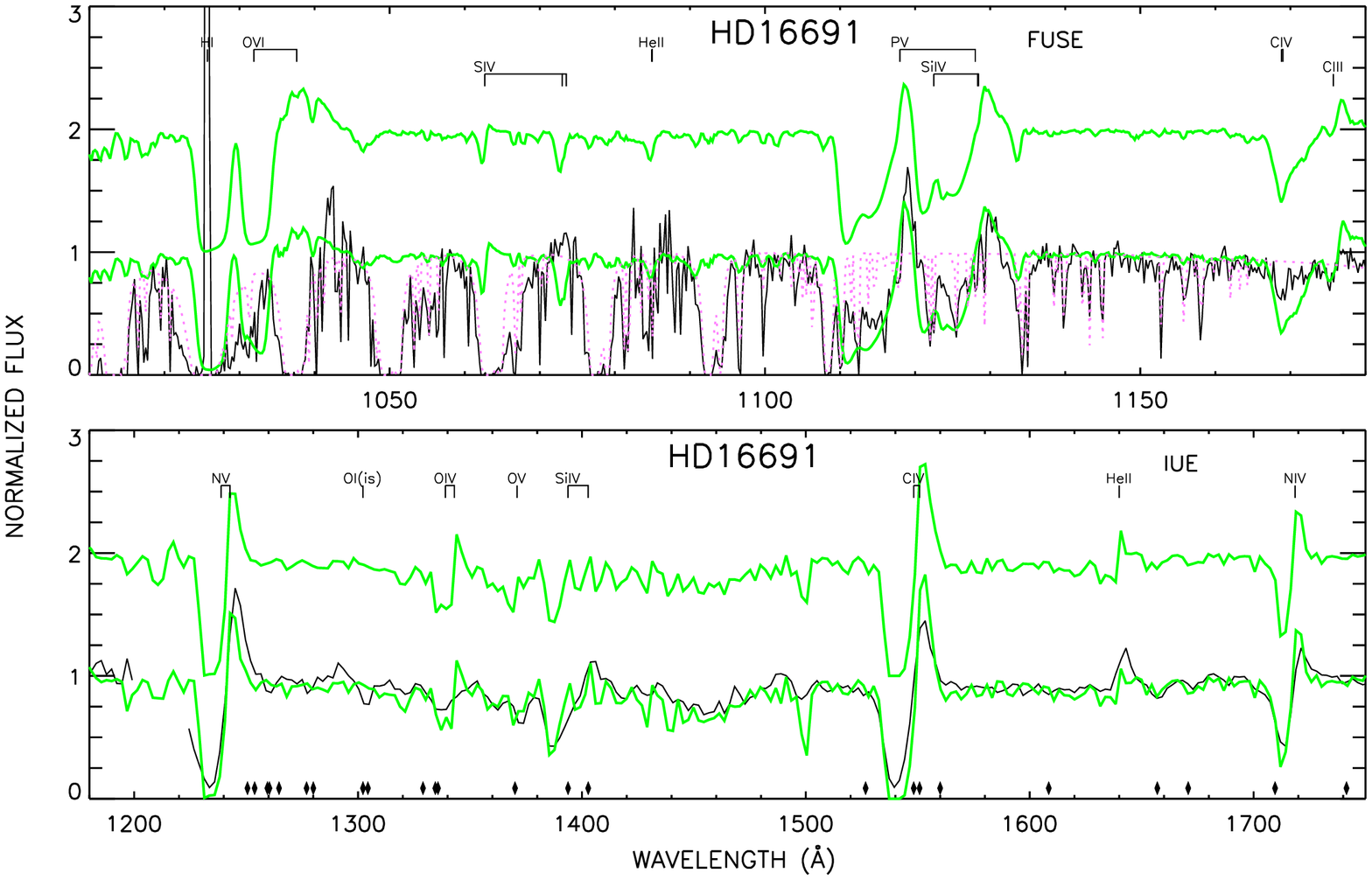}
\includegraphics*[width=8.5cm,angle=90]{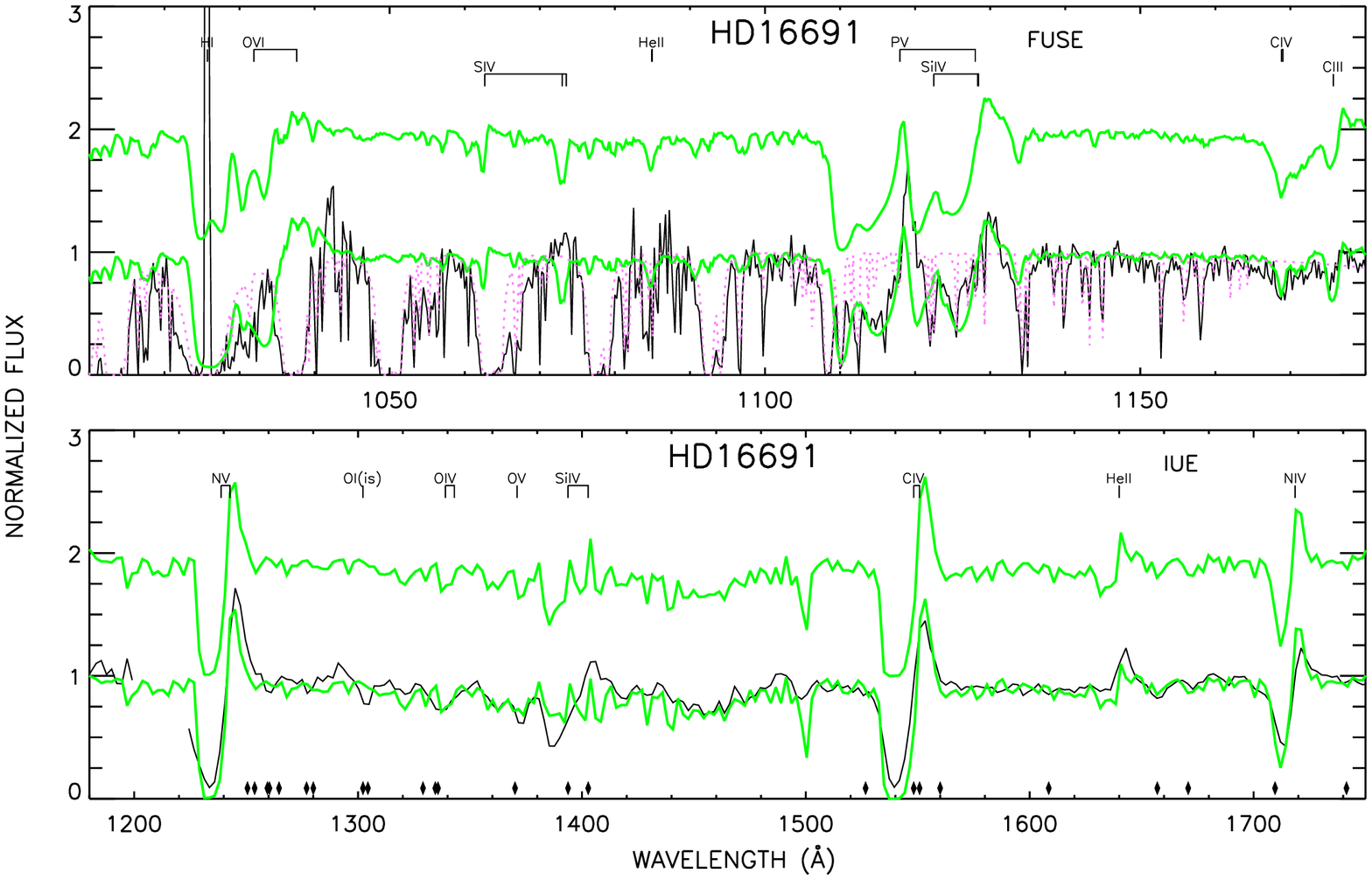}
\end{center}
\caption{HD~16691 spectra and WM-basic models. {\bf Top: }
best-fit models with solar abundances.
The lower model (\Teff=36,000~K, \Mdot=8$\cdot$ 10$^{-6}$\Myr, \Lx=-6.75)
fits better  lines in the IUE range. However,
higher \Teff (upper model, \Teff=38,000~K, \Mdot=1.2$\cdot$ 10$^{-6}$\Myr, \Lx=-6.75)
matches better  the emission component of OVI
and the HeII line.
With the high \Mdot ~necessary to produce 
HeII~$\lambda$1640 emission in the models 
(\Mdot$\rm \ge 7 \cdot 10^{-6}$\Myr, GB04), 
OIV is too strong in any model. This discrepancy can be removed 
adopting  non-solar abundances.
{\bf Bottom:} 
Best-fit models 
with modified CNO abundances: 
$X _{He} = 5 \; X _{He,\odot}$,
$X _C = 0.5 \;  X _{C,\odot}$,
$X _N = 2.0 \;  X _{N,\odot}$,
$X _O = 0.1 \;  X _{O,\odot}$.
The lower model (\Teff=37,000~K, \Mdot=8.3$\cdot$ 10$^{-6}$\Myr, \Lx=-6.75)
fits better NIV, CIV and PV lines but less well the SiIV lines.
The latter are better matched by a model with a lower \Teff (36,000~K) and higher
\Mdot=9.6$\cdot$ 10$^{-6}$\Myr,  
which, however, provides a worse fit to  CIII and PV lines.
The lower oxygen abundance improves the fit to OIV.
}
\end{figure}

\begin{figure}
\label{tefflum}
\begin{center}
\includegraphics[width=10.cm]{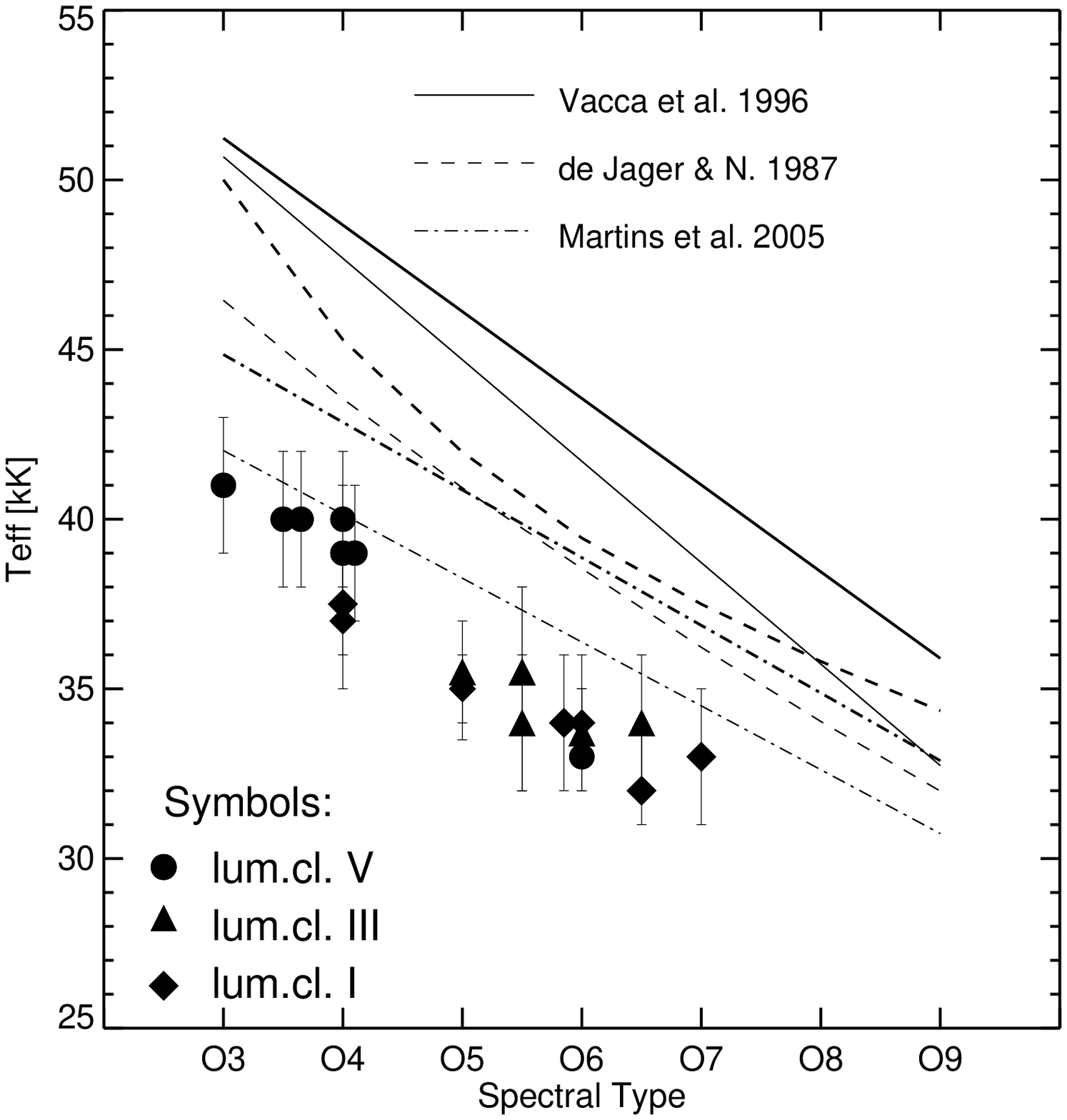}
\end{center}
\caption{
\Teff~ versus spectral type for stars from this paper, BG02, BHG09 and GB04.
Calibrations of
\citet[]{VGS96} (solid line), \citet[]{JN87} (dashed line) and
\citet[]{MSH05} (dash-dotted line) are shown, with thick lines for main  sequence stars and thin
lines for supergiants. 
Although the sample for which both FUSE and high resolution UV spectra are available is very limited, at 
spectral type O4 the difference between luminosity class IV and I in our results is similar
to that seen  in the calibrations by  \citet[]{MSH05} and \citet[]{JN87}, however our derived  
temperatures are lower  across the whole  range. For later types the difference among luminosity classes is
smaller, and the scatter among the data points is higher. 
}
\end{figure}

 Figure~\ref{tefflum} shows the derived \Teff's 
versus spectral type, combining the  samples
from this work, BG02, GB04 and BHG09.
Our results are compared with past calibrations, 
based on compilations from 
plane-parallel, pure H and He analyses 
\citep[]{VGS96,JN87}, and  
with the calibration by 
\citet[]{MSH05},
based on line-blanketed model analysis of optical data. 
The latter is closer to our results, although a small
systematic discrepancy remains.
This discrepancy may  be resolved by combining more diagnostic
lines and with a different treatment 
to constrain wind clumping, as shown by e.g.  BHG09, and by Bouret et al. (2012).   
We stress that this analysis is solely based on UV lines, independent of constraints from 
optical lines, that will be included in a future work (as in BHG09). 
 Such analysis   may 
 provide insight on the characteristics of wind clumping.
Clumping has been discussed e.g. by Puls, Vink \& Najarro (2008), Najarro, 
Hanson, \& Puls, J. (2011),  
see also also Bianchi 2012 and references therein, Herald \& Bianchi 2011, Keller et al. 2011, Kaschinski et al. 2012 for 
discussions on  the importance of a consistent modeling, and on the  
effects on treatement of clumping in UV line analysis in evolved hot stellar
 objects, where similar issues are at play as those recalled in this work
for massive stars, and e.g. Oskinova et al. 2007 for additional discussion. 
Understanding wind clumping is especially 
relevant for reconciling UV and optical line diagnostics, ultimately to  
 precisely constrain mass-loss rate, an important parameter 
to understand  massive stars' evolution. 

~\\
{\bf Acknowledgements: }
We are grateful to the referees for helpful comments. 
This work is based on data obtained 
with the NASA-CNES-CSA {\it FUSE} mission, 
through GI program E085 (P.I. Bianchi), 
and 
 \textit{IUE} archival data obtained from both the INES archive and
the Multimission Archive at the Space Telescope Science Institute (MAST).


\end{document}